\documentclass[english,prd]{article}
\usepackage[latin9]{inputenc}
\usepackage{amsmath}
\usepackage{amssymb}

\makeatletter

\usepackage{pxfonts}

\usepackage{babel}
\makeatother

\begin{document}

\title{On Hawking/Unruh Process: Where does the Radiation Come from?}

\author{Tadas K. Nakamura}

\maketitle
{\hfil CFAAS, Fukui Prefectural University, Fukui 910-1195, Japan
(tadas@fpu.ac.jp)\hfil \par}

\begin{abstract}
The energy source of the radiation in Unruh/Hawking process is investigated
with emphasis on the particle number definition based on conservation
laws. It has been shown that the particle radiation is not the result
of pair creation by the gravitational force, but the result of difference
in the conservation laws to define the particle number. The origin
of the radiated energy in the distant future corresponds to the zero
point oscillations with infinitely large wave numbers. This result
implies the need of reconsideration on the scenario of black hole
evaporation.
\end{abstract}

\section{Introduction}

The theory of quantum particle radiation by gravitational force \cite{hawking,unruh}
is generally accepted as well established, however, several authors
pointed out essential problems that might brow up the whole story
\cite{belinski,helfer,helfer3} (see \cite{helfer2} for a review).
What called trans-Plankian problem has been known from very early
years \cite{gibbons}. In the last two decades, this problem has been
investigated intensively \cite{jacobson,jacobson2}.

The derivation of Hawking/Unruh process is based on so called cis-Plankian
physics, i.e., the theories of gravitation and quantum field we know
at the present. It is believed these theories will break down beyond
an extremely small scale, the Plank scale presumably, and we do not
know what happens there. The unknown physics in that scale is called
trans-Plankian physics. The calculation of Hawking/Unruh effect inevitably
requires the wave modes with infinitesimally small wave length (see,
e.g., \cite{helfer2}), therefore, we need to know the trans-Plankian
physics to understand the radiation mechanism; this is the trans-Plankian
problem. There has been attempts \cite{jacobson2} to derive the radiation
within the cis-Plankian scale, however, they are based on \emph{ad
hoc} assumptions yet to be tested experimentally. 

Helfer \cite{helfer} pointed out another issue we have to consider
before the trans-Plankian problem. The radiation at later times must
have plied up near the horizon at the time of black hole formation,
and its backreaction is far from negligible to the black hole metric.
This implies the theory of Hawking radiation is intrinsically inconsistent
even within the framework of cis-Plankian physics, because such backreaction
may completely destroy the black hole formation. 

This is a serious problem. Papers on the trans-Plankian problem so
far assume that the Hawking radiation is well predicted by the cis-Plankian
physics, and discuss what will happen if we have to consider trans-Plankian
effects. However, if cis-Plankian physics itself fails to derive the
Hawking radiation, then we do not have any reason to believe the existence
of radiation.

The purpose of the present paper is to formulate the Helfer's conclusion
\cite{helfer} from a different point of view. We wish to show the
theory of black hole evaporation is inconsistent even if the cis-Plankian
physics is valid up to infinitesimally small scale. To this end, we
need another unknown physics within the cis-Plankian regime: the effect
of the zero-point oscillation to the gravity. We do not know the general
theory for this, however, there can be two possibilities for the Hawking
radiation. 

One is such that the radiation can carry away the black hole energy
to cause its evaporation. In this case, the backreaction of the quantum
filed is so large as to alter the black hole formation completely
\cite{helfer}. We will see in the present paper there can be another
possibility that the field has no backreaction to the black hole geometry.
In this case, however, there will be no black hole evaporation at
all even with the existence of Hawking radiation. This may be plausible
because there are considerable amount of observational evidence to
believe the existence of black holes.\bigskip{}

We take an approach a little different from the conventional quantization
with creation/annihilation operators in the present paper; the mathematical
structure is equivalent, but its interpretation is not the same. The
quantization with creation/annihilation operators takes two different
steps, transition from the classic to quantum field and introduction
of the particle picture, namely, at the same time. The essential step
in the canonical quantization method is to replace a classical Poisson
bracket with a quantum commutation relation regarding the field as
a collection of quantum operators. If conserved quantities have discrete
eigenvalues with equal intervals, then we can construct the particle
picture. It is well known the latter is not always possible in a curved
spacetime. 

It also should be noted these two steps do not have to be done at
the same place even when we have the particle picture. The commutation
relation must be given on a Chaucy surface on which the Poisson bracket
is defined. The particle picture, in contrast, does not have to be
on the same surface. It can be on some other spacelike surface, which
does not have to be a Chaucy surface as long as there exist some conservation
laws on it. 

In the present paper, the essential quantization, i.e., definition
of the commutation relation, will be done on the surface of constant
time in Minkowski/Kruscal coordinates. Then the particle picture is
introduced based on that quantization, not only on the same surface
but also on the surface of constant time in Rindler/Schwarzschild
coordinates. \medskip{}

The particle picture is based on conservation laws in general. What
we directly measure is not the particle number itself, but some conserved
quantity such as energy or electric charge. We imagine there are $n$
particles, each of which carries a certain amount of conserved quantity,
if the total of the quantity has discrete values proportional to an
integer $n$.

This means the concept of {}``particle number'' is defined by conserved
quantities. If all the conserved quantities share the same $n$, then
we can define one unique particle number, however, this is not the
case. There can be several different definitions of particle numbers
because there can be several different Killing vector fields that
determine the conservation laws in a relativistic spacetime.

Consequently one physical state can have different particle numbers,
and this is what is happening in the Unruh/Hawking process. Particles
are not created in the literal meaning of {}``creation'', which
means the particle number increases as time goes on. Rather, what
takes place is just a difference of particle numbers caused by the
difference in their definitions. This is in agreement with the result
of Belinski \cite{belinski} calculated from another viewpoint.

We will see in the present paper the radiation of particles comes
from the vacuum state, i.e., zero particle state, of another kind
of particle number. This is possible because a vacuum is not a completely
empty space but has zero point oscillations. The continuous particle
radiation can take place because the zero point oscillations exist
up to infinitely large wave numbers, which means infinite amount of
energy source.

\bigskip{}

The present paper is organized in the following. In section 2 we first
review some basic concepts to clarify the procedure of quantization
used in the later sections. We examine in Section 3 the case of Unruh
process in a flat spacetime, since it has the two different types
of conservation laws clearly defined; we can understand the problem
with this simple analogy. We apply the results obtained in Section
3 to the case of Schwarzschild black holes in Section 4, and a brief
summary is given in Section 5.

\section{Basics}

\subsection{Time and Energy}

The concept of energy is often used in a sloppy way, which sometimes
leads to misconceptions. The integration of the energy-momentum tensor
cannot be carried out in the curved spacetime in general, however,
there can be well defined {}``energy'' as a globally conserved quantity
if there exists a Killing vector field. If a Killing vector $\xi_{\nu}$
is timelike, then the integration $\int_{\Sigma}\xi_{\nu}T^{\nu\mu}d\Sigma_{\mu}$
($T^{\nu\mu}$: energy-momentum tensor) over an appropriate spacelike
surface $\Sigma$ is conserved with respect to the time evolution
in $\xi_{\nu}$. If there are several different timelike Killing vector
fields, there can be the same number of corresponding energies; the
energies defined by different Killing vector fields are different
physical entities.

Sometimes this difference in energies is not well understood and causes
confusion; one good example is an intuitive explanation of the Hawking
radiation found in popular science books. It goes like: (1) a vacuum
is not an empty space but filled with instantaneous pair production
of virtual particles; (2) a pair of virtual particles can exist within
a short time period of $\Delta t\sim\hbar/\Delta E$ because of the
uncertain principle; (3) when such virtual particles are created near
the event horizon, one of the pair may fall into the black hole across
the horizon during the time interval of $\Delta t$; (4) once a virtual
particle crosses the horizon, its energy becomes negative; (5) then
the other particle of the pair can have positive energy without violating
the energy conservation law.

This explanation does not specify the Killing vector field with which
the time and energy are defined. If the Killing vector is something
like the Schwarzschild time, then a particle takes infinitely long
time to reach the horizon, and cannot cross the horizon during the
period of $\Delta t$. If, on the other hand, the Killing vector is
such that a particle can cross the horizon within a finite period,
then the corresponding energy does not change the sign on the other
side of the horizon. The pair production near the horizon is not likely
to occur to cause the Hawking radiation in both cases.

\subsection{Conservation Laws and Particle Numbers}

Usually the quantization process to investigate Hawking/Unruh process
is based on the creation/annihilation operators defined by the negative/positive
frequency modes. In the present paper we take one step backwards and
perform the quantization by replacing the Poisson bracket with the
commutation relation. Hereafter, let us use the word {}``quantization''
with the meaning of the transition from the classical to quantum theory,
and does not necessarily mean the particle picture. 

The particle picture is derived from conserved quantities after the
quantization. If the quantum observable of a conserved quantity has
the structure of a harmonic oscillator, its eigenvalues are proportional
to $n+\frac{1}{2}$ with $n=0,1,2,\cdots$. Usually the constant $\frac{1}{2}$
is subtracted out by normal ordering, thus the quantity is proportional
to $n$. When there are several conserved quantities that share the
same $n$ for the same state, then we can interpret $n$ as the particle
number.

Suppose we establish quantization somehow, and find an observable
$\hat{a}$ (hat mark indicates a quantum operator) and its Hermite
conjugate $\hat{a}^{\dagger}$ have the following commutation relation
\begin{equation}
[\hat{a},\hat{a}^{\dagger}]=\hat{a}\,\hat{a}^{\dagger}-\hat{a}^{\dagger}\hat{a}=1\,.\label{eq:commutation1}\end{equation}
We use the unit system with $G=\hbar=c=1$ throughout the present
paper. The general theory of quantum harmonic oscillators tells us
(see., e.g., \cite{messiah}) an observable defined as \begin{equation}
\hat{A}=\frac{A_{0}}{2}\left(\hat{a}\,\hat{a}^{\dagger}+\hat{a}^{\dagger}\hat{a}\right)\end{equation}
has the eigenstates $\left|n_{A}\right\rangle $ that satisfies\begin{equation}
\hat{A}\left|n_{A}\right\rangle =A_{0}\left(n+\frac{1}{2}\right)\left|n_{A}\right\rangle \;,(n=0,1,2\cdots,)\,.\label{eq:eigen}\end{equation}
if $\hat{a}$ has the commutation relation of  (\ref{eq:commutation1}). 

For the above argument the observable $\hat{A}$ does not have to
be related to the Hamiltonian explicitly (note: the Poisson bracket
has something to do with the Hamiltonian implicitly), and there can
be several choices for such observables. For example if we define
a new observable $\hat{b}$ as\begin{equation}
\hat{b}=\alpha\,\hat{a}+\beta\,\hat{a}^{\dagger}\end{equation}
with $\alpha^{2}-\beta^{2}=1$ then it also satisfies the commutation
relation like  (\ref{eq:commutation1}) and thus $\hat{B}=\frac{1}{2}B_{0}(\hat{b}\,\hat{b}^{\dagger}+\hat{b}^{\dagger}\hat{b})$
has the eigenvalues $B_{0}(n+\frac{1}{2})$. It is easy to confirm
its eigenstates $\left|n_{B}\right\rangle $ are not the eigenstates
of $\hat{A}$, i.e., $\hat{A}\left|n_{B}\right\rangle \ne(n+\frac{1}{2})\left|n_{B}\right\rangle $,
and vice versa.

Both pairs $\hat{a}$, $\hat{a}^{\dagger}$ and $\hat{b}$, $\hat{b}^{\dagger}$
have the structure of annihilation/creation operators, however, it
is not enough for the particle picture. To construct the particle
picture with $\hat{a}$ and $\hat{a}^{\dagger}$, $\hat{A}$ must
obey a conservation law in time, i.e., $\partial\hat{A}/\partial t=0$
(here we employ the Heisenberg picture) at least approximately. If
$\hat{A}$ rapidly changes even without interactions, so does $n$,
and it is not appropriate to regard $n$ as a particle number.

\bigskip{}

When the Hamiltonian does not depend on time explicitly, the condition
of $\partial\hat{A}/\partial t=0$ is equivalent to the following
commutation relation: \begin{equation}
[\hat{A},\hat{H}]=0\,.\label{eq:conserve}\end{equation}
Obviously the Hamiltonian itself satisfies this condition, therefore,
the Hamiltonian is usually used to introduce the particle picture.
Then the operators $\hat{a}$ and $\hat{a}^{\dagger}$ become the
amplitudes of wave modes with positive and negative frequencies respectively,
which are usually used in the procedure of quantization as the annihilation
and creation operators. Therefore, the mathematical structure in the
present paper is equivalent to the one in the conventional method.

If there are other conserved observables with respect to the time
$t$, then they share the same set of eigenstates with the Hamiltonian
$\hat{H}$ because of the commutation relation  (\ref{eq:conserve}).
Therefore $n$ can be regarded as the particle number without specifying
the conserved quantities, as long as the conservation laws are on
the same time evolution of $t$. Especially, the ground sate of the
observables is uniquely determined, and we call it {}``vacuum''.
We can define the number operator as\begin{equation}
\hat{N}=\hat{a}^{\dagger}\hat{a},\label{eq:number}\end{equation}
whose eigenvalue is the particle number $n$, and the vacuum means
the eigenstate with $n=0$.

However, in relativistic spacetimes there can be several different
types of time evolution with different sets of conservation laws because
several different Killing vector fields can exist; the Minkowski and
Rindler times in a flat spacetime are a good example. 

If two different types of time evolution have their own conservation
laws, then the conserved quantities that belong to different time
evolution may have different sets of eigenstates. Consequently the
ground state in one time evolution is not the ground state in another,
in other words, they have different vacuum states. This is what causes
Hawking/Unruh radiation as we will see in the next section.

\section{Unruh process}

\subsection{Minkowski Coordinates}

Suppose the following real valued Klein-Goldon equation in a two dimensional
Minkowski spacetime where $t$ and $x$ are the time and space coordinates:\begin{equation}
\phi_{,tt}-\phi_{,xx}=0\,.\label{eq:KleinGoldon}\end{equation}
We write $\partial\phi/\partial t=\phi_{,t}$ etc.\ in shorthand.
We take the Cauchy surface for the canonical dynamics as the one defined
with $t=\textnormal{constant}$, then the Hamiltonian may be written
as \begin{equation}
H=\int_{-\infty}^{\infty}\frac{1}{2}\left[\phi_{,t}^{2}(t,x)+\phi_{,x}^{2}(t,x)\right]\, dx\,.\label{eq:hamilton}\end{equation}
We expand the field as \begin{equation}
\phi(x,t)=\int\left[a(k)\, u(k;x,t)+a^{*}(k)\, u^{*}(k;x,t)\right]dk\,,\label{eq:expansion}\end{equation}
with mode functions \begin{equation}
u(k;x,t)=\frac{1}{\sqrt{4\pi\omega}}\, e^{-i\omega t+ikx}\,,\end{equation}
where $\omega=|k|$ and the asterisk indicates complex conjugate.
Precisely speaking, $a_{k}$ diverges to infinity as we usually encounter
in the Fourier transform; we assume some appropriate prescription,
such as the distribution/hyperfunction formulation, has been applied
to avoid this difficulty in this paper.

The essential transition from the classical to the quantum field is
done by replacing $a(k)$ and $a^{*}(k)$ with the operators satisfying
the commutation relation of\begin{equation}
\left[\hat{a}(k),\hat{a}(k')^{\dagger}\right]=\hat{a}(k)\,\hat{a}^{\dagger}(k')-\hat{a}^{\dagger}(k)\,\hat{a}(k')=\delta(k-k')\,.\label{eq:commutation2}\end{equation}
The above quantization is based on the Cauchy surface of $t=\textnormal{constant}$.
It should be noted that the quantization in this paper takes place
only once at this point. Later we introduce the particle picture on
the surface of constant Rindler time, but it is expressed by a linear
superposition of $\hat{a}$ and $\hat{a}^{\dagger}$ and based on
the commutation relation defined here. 

The Hamiltonian  (\ref{eq:hamilton}) can be expressed as a collection
of quantum harmonic oscillators:\begin{equation}
\hat{H}=\int\frac{\omega}{2}\left[\hat{a}(k)\,\hat{a}(k)^{\dagger}+\hat{a}^{\dagger}(k)\,\hat{a}(k)\right]dk\,.\end{equation}
Therefore, we can define the particle number as explained in the previous
section, and ground sate of $\hat{H}$ is called {}``vacuum''.

Now that we have the quantized operators $\hat{a}(k)$ and $\hat{a}^{\dagger}(k)$,
we can calculate the field $\hat{\phi}(x,t)$ at any point of the
spacetime as a quantum observable. Any classical quantity defined
from the classical field $\phi$ can be quantized by replacing \begin{equation}
\phi\rightarrow\hat{\phi}=\int\left[\hat{a}(k)\, u(k;x,t)+\hat{a}^{\dagger}(k)\, u^{*}(k;x,t)\right]dk\,.\end{equation}

\subsection{Rindler Coordinates}

The Hamiltonian $H$ in  (\ref{eq:hamilton}) is the energy with the
conservation law based the Killing vector field of Minkowski time
$\partial_{t}$. We examine in the following another conservation
law resulting from another Killing vector field $\kappa x\partial_{t}-\kappa t\partial_{x}$,
where $\kappa$ is a real constant that corresponds to the relativistic
acceleration. The energy $M$ for this Killing vector field is written
in the classical field theory as 

\begin{equation}
M=\int_{\Sigma(\eta)}\frac{1}{\kappa(t^{2}-x^{2})}\left(t\phi_{x}^{2}+x\phi_{,t}^{2}\right)d\Sigma\,,\end{equation}
where $\Sigma(\eta)$ is a surface specified by $t/x=\tanh(\kappa\eta)$
and $x>0$. Clearly $M$ is not the same quantity as $H$, therefor,
let us distinguish $M$ and $H$ by calling them {}``Rindler energy''
and {}``Minkowski energy'' respectively.

The density of the Rindler energy is conserved locally, and there
is no Rindler energy flow across the left and right Rindler wedges,
thus we have $\partial M/\partial\eta=0$; once we calculate $M$
on a surface $\Sigma(\eta)$ with a given $\eta$, then the result
holds for all $\eta$. When we choose $\eta=0$, we can express $M$
with the coefficients $a(k)$ of the Minkowski modes. Since $M$ is
quadratic in $\phi$ we can write \begin{equation}
M=\iint\left[A(k,k')\, a(k)\, a(k')+B(k,k')\, a(k)\, a^{*}(k')+C(k,k')\, a^{*}(k)\, a^{*}(k')\right]dk\, dk'\,,\label{eq:defM}\end{equation}
with coefficients $A$, $B$, and $C$ that do not depend on $\eta$
or $t$.

The above quantity is quantized by replacing $a(k)\rightarrow\hat{a}(k)$
and $a^{*}(k)\rightarrow\hat{a}^{\dagger}(k)$ as \begin{equation}
\hat{M}=\iint\left[A\,\hat{a}(k)\,\hat{a}(k')+\frac{B}{2}\left(\hat{a}(k)\,\hat{a}^{\dagger}(k')+\hat{a}^{\dagger}(k)\,\hat{a}(k')\right)+C\,\hat{a}^{\dagger}(k)\,\hat{a}^{\dagger}(k')\right]dk\, dk'\,.\label{eq:quantumM}\end{equation}
As noted before, this quantization is based on the Cauchy surface
of $t=\textnormal{constant}$, not $\Sigma(\eta)$. Therefore, $\Sigma(\eta)$
does not have to be a Cauchy surface.\bigskip{}

What we wish to show in the following is that the particle numbers
defined by $\hat{M}$ and $\hat{H}$ are not the same. Before that,
we have to show that $\hat{M}$ surely can define the particle number.
Suppose an operator $\hat{b}$ is defined as a linear superposition
of $\hat{a}$ and $\hat{a}^{\dagger}$ as\begin{equation}
\hat{b}(p)=\int\left[\alpha(p,k)\,\hat{a}(k)+\beta(p,k)\,\hat{a}^{\dagger}(k)\right]dk\,.\label{eq:bogolubov}\end{equation}
If $\hat{b}$ satisfies the commutation relation\begin{equation}
[\hat{b}(p),\hat{b}^{\dagger}(p)]=\delta(p-p')\,,\label{eq:commutation3}\end{equation}
and $\hat{M}$ can be expressed with $\hat{b}$ as\begin{equation}
\hat{M}=\frac{1}{2}\int\sigma\left[\hat{b}(p)\,\hat{b}^{\dagger}(p)+\hat{b}^{\dagger}(p)\,\hat{b}(p)\right]dp\,,\label{eq:comM}\end{equation}
then we can define the particle number with $\hat{M}$ in the similar
way as done in the previous subsection with $\hat{H}$.

It is possible show the above two equations with direct calculation,
however, it is easier to use the following Rindler coordinates $(\eta,\rho)$
for the right Rindler wedge i.e., region of $x>0,$ \textbar{}x\textbar{}\textgreater{}\textbar{}t\textbar{}:
\begin{equation}
t=\rho\sinh(\kappa\eta),\; x=\rho\cosh(\kappa\eta)\,.\end{equation}
Then the Rindler energy $\hat{M}$ may be written as\begin{equation}
\hat{M}=\int_{0}^{\infty}\frac{1}{\kappa\rho}\left(\hat{\phi}_{,\eta}^{2}+\kappa^{2}\rho^{2}\hat{\phi}_{,\rho}^{2}\right)d\rho\,.\label{eq:rindM}\end{equation}
We introduce the eigenfunctions\begin{equation}
v(p;\eta,\rho)=\frac{1}{\sqrt{4\pi\sigma}}\,\exp(-i\sigma\eta+i\kappa^{-1}p\ln(\kappa\rho))\,,\label{eq:defv}\end{equation}
with $\sigma=|p|$. The wave function $\hat{\phi}$ can be expanded
in the right Rindler wedge as\begin{equation}
\hat{\phi}=\int\left[\hat{b}(p)\, v(p)+\hat{b}^{\dagger}(p)\, v^{*}(p)\right]dp\,,\end{equation}
then $\hat{M}$ in  (\ref{eq:rindM}) can be cast into  (\ref{eq:comM}).
In this context $\alpha$ and $\beta$ in  (\ref{eq:bogolubov}) are
equivalent to the Bogolubov coefficients that satisfy \begin{equation}
\int\left[\alpha(p_{1},k)\alpha^{*}(p_{2},k)-\beta(p_{1},k)\beta^{*}(p_{2},k)\right]dk=\delta(p_{1}-p_{2})\,,\label{eq:alphabeta}\end{equation}
 The above property combined with  (\ref{eq:commutation2}) yields
the commutation relation of  (\ref{eq:commutation3}), therefore we
can define particle numbers with $\hat{M}$.\bigskip{}

What we do next is to compare the eigenstates of $\hat{M}$ and $\hat{H}$.
From  (\ref{eq:quantumM}) it is clear that $\hat{M}$ and $\hat{H}$
do not share the same set of eigenstates unless $A$ and $C$ vanishes.
We can see from (\ref{eq:bogolubov}) and (\ref{eq:defM}) that $A$
and $C$ vanished when $\beta=0$, however, $\beta$ can be evaluated
by a straightforward integration (see, e.g., \cite{dewitt}), resulting\begin{equation}
\left|\beta(p,k)\right|^{2}=\frac{1}{2\pi\omega}\,\frac{1}{e^{2\pi\sigma/\kappa}-1}\ne0\,.\label{eq:beta}\end{equation}
Therefore, the eigenstates of $\hat{M}$ are not the eigenstates of
$\hat{H}$. This means that the ground state of $\hat{H}$ is not
the ground state of $\hat{M}$, which can be stated in other words
as {}``a vacuum defined by the Minkowski energy is not a vacuum defined
by the Rindler energy''. The expected value of the particle number
defined by Rindler energy (we call Rindler particle number hereafter)
in the Minkowski vacuum can be calculated using  (\ref{eq:beta}),
resulting the well known Plankian distribution \cite{dewitt}. 

What we have seen above is not surprising since different operators
may have different sets of eigenstates. However, it contradicts with
the picture of the pair production by the gravitational force often
found in intuitive explanations like the one in Section 2. The particle
number defined by Minkowski energy (Minkowski particle number hereafter)
is zero for all $t$, and the Rindler particle number has the fixed
Plankian distribution for all $\eta$. This is consistent with the
time symmetry; the vacuum in the flat spacetime must be symmetric
in time, both in $t$ and $\eta$, but the particle creation process
is not symmetric.

\subsection{Origin of the Radiation}

The Rindler particle number in the Unruh process is calculated by
the expected value of the Rindler energy for the ground state of the
Minkowski energy, i.e., $\left\langle 0_{H}\right|\hat{M}\left|0_{H}\right\rangle $.
If the {}``vacuum'' were a completely empty space, i.e., $\hat{a}\left|0_{H}\right\rangle =\hat{a}^{\dagger}\left|0_{H}\right\rangle =0$,
then $\left\langle 0_{H}\right|\hat{M}\left|0_{H}\right\rangle $
would vanish. However, this is not true since the quantum ground state
has zero point oscillation, and hence $\left\langle 0_{H}\right|\hat{M}\left|0_{H}\right\rangle \ne0$.
This means the {}``particles'' found in the Unruh process comes
from the zero point oscillation of the Minkowski modes. 

Then what we wish to know is the properties of zero point oscillations
that contribute to the continuous radiation of Rindler energy. In
the present paper we concentrate on right moving waves, i.e., $\omega k<0$
or $\sigma p<0$, since their analog in the Schwarzschild spacetime
play the key role in the black hole evaporation. It should be noted,
however, left moving waves are also problematic and should be examined
in the next step. We first examine the properties of waves in the
classical limit, and then apply the result to the quantum vacua. 

To begin with, we observe that the a eigenmode  (\ref{eq:defv}) has
infinite Rindler energy in a finite region of $0\le\rho<\rho_{c}$
with arbitrary position $\rho_{c}$ in the right Rindler wedge; this
can be confirmed by the following direct integration:\begin{equation}
\lim_{\varepsilon\rightarrow0}\int_{\varepsilon}^{\rho_{c}}\frac{1}{\kappa\rho}\,\left(\hat{v}_{,\eta}^{2}+\kappa^{2}\rho^{2}\hat{v}_{,\rho}^{2}\right)\, d\rho\rightarrow\infty\,.\end{equation}
Also it is easy to confirm there is a constant rightward outflux of
the Rindler energy at $\rho=\rho_{c}$ by direct calculation. This
outflux comes from the region of $0\le\rho<\rho_{c}$, but the total
Rindler energy can be conserved because the amount of the Rindler
energy in that region is infinitely large. Belinski \cite{belinski}
considered this fact as physically unacceptable and concluded the
radiation results from $v(p)$ is just a mathematical illusion. 

The present paper takes a different interpretation. The infinite Rindler
energy can be physically real as long as we believe zero point oscillations
exist for any high frequency modes, because the collection of such
oscillations has infinitely large Rindler energy even in a finite
volume. 

To see this, we examine the behavior of the a wave packet in the following
form ({}``$c.c.$'' means complex conjugate):\begin{equation}
\phi(\eta,\rho)=\exp\left(\frac{-(\rho-e^{\kappa\eta}\,\rho_{0})^{2}}{(e^{\kappa\eta}\, s_{0})^{2}}\right)\,\exp[-i\sigma\eta+i\kappa^{-1}p\ln(\kappa\rho)]+c.c.\,.\end{equation}
This wave packet was initially localized around $\rho=\rho_{0}$ with
width $s_{0}$ at $\eta=0$, and propagates rightward. The width of
the packet becomes larger and the wave number becomes smaller as a
result of wave propagation.

The wave packet can be expanded by the Minkowski modes $u(k;t,x)$
as\begin{equation}
\phi(t,x)=\int\left[(\phi,u(k))\, u(k;t,x)+(\phi,u^{*}(k))\, u^{*}(k;t,x)\right]\, dk\label{eq:expansion2}\end{equation}
with the Klein-Goldon inner products $(\phi_{1},\phi_{2})$, which
can be calculated at $t=\eta=0$ as\begin{equation}
(\phi_{1},\phi_{2})=\int\left[\phi_{1,t}\phi_{2}^{*}-\phi_{1}\phi_{2,t}^{*}\right]_{t=0}\, dx\,.\label{eq:KGproduct}\end{equation}
When $s\ll\rho_{0}\ln(\kappa\rho_{0})$ then we can approximate \begin{equation}
\phi(\eta,\rho)\simeq\exp\left(\frac{-(\rho-\rho_{1})^{2}}{s_{1}^{2}}-ip\eta+\frac{ip}{\kappa}\left[\ln(\kappa\rho_{1})+\frac{1}{\rho_{1}}\,(\rho-\rho_{1})\right]\right)+c.c.\,,\label{eq:approxpacket}\end{equation}
where $s_{1}=s_{0}\, e^{\eta}$ and $\rho_{1}=\rho_{0}\, e^{\eta}$
are the width and center of the wave packet at a time $\eta$. Using
the above approximation we obtain\begin{equation}
\phi(t,x)=\frac{2}{s_{0}\sqrt{\pi}}\exp\left[i\kappa^{-1}p\ln(\kappa\rho_{0})\right]\int e^{-ik\rho_{0}}\exp\left[-s_{0}^{2}(k-p/\kappa\rho_{0})^{2}\right]e^{-i\omega t+ikx}dk+c.c.\end{equation}
from  (\ref{eq:expansion2}) with  (\ref{eq:KGproduct}). The above
expression means that the wave packet comes from the Minkowski modes
with wave number around $k_{0}=p/\kappa\rho_{0}$ when $s_{0}\gg\kappa\rho_{0}/p$.

Suppose we find a wave packet around $\rho_{1}$ at a given time $\eta=\eta_{1}\,(>0)$
in the Rindler space then its position at $\eta=0$ was $\rho_{0}=\rho_{1}\, e^{-\kappa\eta_{1}}$,
therefore, the packet consists of the Minkowski modes with $k\sim k_{0}=p\, e^{\kappa\eta_{1}}/\kappa\rho_{0}$.
When we regard the wave field at $\eta=\eta_{1}$ as a superposition
of such wave packets, we see that the waves in the region of $0<\rho<\rho_{1}$
at $\eta=\eta_{1}$ consists of Minkowski modes with wave numbers
larger than $k_{0}=p\, e^{\kappa\eta_{1}}/\kappa\rho_{0}$. 

Since $k_{0}\rightarrow\infty$ in the limit of $\eta_{1}\rightarrow\infty$,
we understand the Rindler energy radiation at the distant future in
$\eta$ comes from the Minkowski modes with infinitely large wave
numbers. The Rindler coordinates represent an observer with constant
acceleration, and the relative velocity of the accelerating observer
to the rest frame becomes infinitely large in the limit of $\eta\rightarrow\infty$.
Waves with finite wave numbers in this limit are infinitely red shifted,
therefore its original wave number must have been infinitely large.

\bigskip{}

Now let us apply the above observation to quantum vacua to see the
origin of the Rindler particles. Suppose the quantum state is Minkowski
vacuum, i.e., the ground state of the Minkowski energy. Then the state
has zero point oscillation up to infinitely large wave numbers. Usually
the energy of these zero point oscillation is subtracted out by normal
ordering, and we regard there is no particle in the ground state.
However, the ground state of the Minkowski energy is not the ground
state of the Rindler energy, which means the existence of the Rindler
particles. The continuous radiation of Rindler particles is possible
for any large $\eta$ because the zero point Minkowski energy exist
for modes with any large wave numbers. The radiated Rindler energy
must have been piled up near $\rho=0$ at the initial time of $\eta=0$,
since there is no particle creation as we have seen in the previous
subsection.

\section{Hawking Radiation}

Let us move on to the Hawking radiation from a Schwarzchild black
hole in this section. We introduce the Schwarzchild coordinates $(t,r,\theta,\varphi)$
whose metric is \begin{equation}
ds^{2}=\left(1-\frac{2M}{r}\right)dt^{2}-\left(1-\frac{2M}{r}\right)^{-1}dr^{2}-r^{2}d\theta^{2}-r^{2}\cos^{2}\theta d\varphi\,.\end{equation}
The Kruscal coordinates $(u,v,\theta,\varphi)$ are related to $(t,r,\theta,\varphi)$
as\begin{eqnarray}
u^{2}-v^{2} & = & 2M(2M-r)\exp\left(\frac{r}{2M}\right)\,,\\
\left|\frac{u-v}{u+v}\right| & = & \exp\left(\frac{t}{2M}\right)\,.\end{eqnarray}
 The rest of the coordinates, $\theta$ and $\varphi$, are unchanged. 

It is generally accepted that the quantum properties of vacuum near
the Schwarzchild event horizon is essentially the same as those in
the Rindler spacetime \cite{unruh,fulling}, therefore, the results
we obtained in the previous section are basically valid by replacing
Minkowski/Rindler coordinates with Kruscal/Schwarzchild coordinates
(note: $t$ in the Schwarzchild spacetime corresponds to $\eta$,
not $t$, in the flat spacetime). There are, however, two fundamental
differences. One is the definition of the energy in Kruscal coordinates,
and the other is the backreaction of the quantum fields to the black
hole metric. The latter causes the essential problem in the scenario
of the black hole evaporation.\bigskip{}

The first difference is about the energy that corresponds to the Minkowski
energy. The Kruscal time $u$ is not a global Killing time, and thus
there is no global energy conservation law like for the Minkowski
energy. However, $u$ can be approximately regarded as a Killing time
near the horizons. As we have seen in the previous section, the radiation
in later time comes from the infinitely high frequency modes infinitesimally
near the horizon, therefore the energy is conserved approximately
for these waves. 

This is in parallel to the approximation of geometrical optics used
by Hawking in his original paper \cite{hawking}; geometrical optics
assumes locally constant frequency, which means locally constant energy.
Hereafter we assume the energy corresponds to the Kruscal time $u$
is approximately conserved, and treat it in the same way as for the
Minkowski energy in the previous subsection. We call it Hartle-Hawking
energy since its ground state is often called Hartle-Hawking vacuum.
The energy defined with the Schwarzchild time is called Boulware energy
hereafter for the same reason.

We have another problem in the definition of the Hartle-Hawking energy
in a Schwarzchild spacetime. Minkowski energy is defined as an integration
over a surface of $t=\textnormal{constant}$ in a flat spacetime.
If we introduce a similar definition for the Hartle-Hawking energy
with Kruscal time $u=\textnormal{constant}$, the energy would include
the part of the white hole in the extended Schwarzchild spacetime.
This difficulty may be avoided by analyzing the black hole formation
process by a star collapse, or the analytical continuation method
proposed by Hartle and Hawking \cite{hartle}. A detailed analysis
on this point will be given in a forthcoming paper of the author.
\bigskip{}

The second difference is far more serious; the energy of zero point
oscillations may change the metric. Usually the zero point energy
is subtracted out by normal ordering in the source term of the Einstein
equation, and the vacuum does not have effect on the metric. This
means only the excited state of the energy can cause gravitation.
However, as we have seen in the previous section, the ground state
of Hartle-Hawking energy is not the ground state of the Boulware energy,
and vise versa. 

In a flat spacetime we consider the ground state of the Minkowski
energy is the state of no gravitation, because the Minkowski coordinates
are the {}``natural'' coordinate system. We have seen there is infinite
accumulation of the Rindler energy near $\rho=0$ for a Rindler mode
$v(p;\eta,\rho)$. There must be an infinitely strong source of gravitational
force at $\rho=0$ if we assume the ground state of the Rindler energy
is the state of no gravitation, since the vacuum state defined by
the Minkowski energy is the excited state of the Rindler energy. This
is not plausible, and we can conclude the ground state of the Minkowski
energy has no effect on the metric.

In contrast, we do not know which coordinate system is {}``natural''
to calculate the energy (stress-energy tensor) for a curved spacetime
in general (see, e.g., \cite{birrel-fulling}). The Schwarzchild coordinates
are implicitly assumed to be {}``natural'' in the scenario of the
black hole evaporation, in other words, excited states of Boulware
energy causes the gravity. The black hole evaporation is believed
to be the result of the Hawking radiation that carries the energy
away from the black hole, and the energy in this context is the Boulware
energy; this means the Boulware energy can have effects on the black
hole metric somehow. 

If this is true, however, the Rindler energy radiated at later times
must have been exist just outside of the horizon from the beginning
\cite{belinski}. The energy does not come from inside the black hole,
but comes from the Minkowski modes with extremely high wave numbers.
This means the backreaction of the quantum field is far from negligible
to the black hole metric \cite{helfer}.

On the contrary, we can imagine the gravity is caused by the exited
state of Hartle-Hawking energy and its ground state has no effect
on the metric. The black hole can exist in this case, however, it
cannot evaporate. There exists a constant outflow of Boulware energy,
but it is the ground state of the Hartle-Hawking energy and does not
have a backreaction on the metric. Consequently the black hole metric
is unchanged at all, just like the Unruh process does not change the
flat metric. There can be other possibilities for the effect of the
zero point energy to the metric, however, it is hard to imagine there
is an extremely convenient case which is favorable for the scenario
of black hole evaporation.

\section{Summary}

What we have seen in the present paper are:

\begin{enumerate}
\item The ground state of Minkowski/Hartle-Hawking energy is not the ground
states of Rindler/Boulware energy, and this is what causes the Hawking/Unruh
process.
\item The quantum state is unchanged and particles pairs are not created
in any coordinate system; the number of Minkowski/Hartle-Hawking particles
is zero and number of Rindler/Boulware particles has time stationary
Plankian distribution all through the time, where {}``time'' means
the Rindler/Schwarzchild time .
\item The radiation of Rindler/Boulware energy in the distant future of
Rindler/Schwarzchild time comes from the zero point oscillation of
Minkowski/Hartle-Hawking energy with infinitely large wave frequencies.
\item The effect of zero point energy to the metric is not known, however,
we have the following two possibilities for a Schwarzchild spacetime.
The scenario of black hole evaporation is inconsistent in both cases.

\begin{enumerate}
\item If the Hawking radiation causes the black hole evaporation, it means
the excitation in the Boulware energy can cause the metric change.
The Boulware energy radiated later time was accumulated near the horizon
at the initial time, whose existence essentially alter the Schwarzchild
metric from the beginning.
\item If, on the contrary, the Boulware energy of the Hawking radiation
does not affect the metric, then the Schwarzchild metric can exist
as we expect, but exists forever. There is no evaporation of the black
hole.
\end{enumerate}
\end{enumerate}
We started the present study by assuming that the cis-Plankian physics
is valid for any small scale phenomena, and end up with the inconsistency
of black hole evaporation. This fact means we have no reason to believe
the black hole evaporation. A new theory of physics in trans-Plankian
scale may save the evaporation, but may not; we can imagine anything,
but cannot believe. What we can say for sure is that the physics we
know at the present is not able to predict the black hole evaporation,
if the calculations in the present paper are correct.

We see the scenario of black hole evaporation is inconsistent, however,
we do not know what is the consistent theory even within the cis-Plankian
regime. The problem deeply depends on the renormalization procedure
in curved spacetimes, to which we do not know the answer yet. It is
often said Hawking process can be a touchstone for the theory of quantum
gravity. The author of the present paper would like to say it also
can be a touchstone for the renormalization theory, or theory on what
is avoided by renormalization at the present, in curved spacetimes.


\begin{thebibliography}{10}
\bibitem{hawking}Hawking, S. W., Comm. Math. Phys. \textbf{43}, 199
(1975).

\bibitem{unruh}Unruh, W. G., Phys. Rev.\textbf{ D14}, 870 (1976).

\bibitem{belinski}Belinski, V. A., Phys. Lett. \textbf{A209}, 13
(1995); Phys Lett. \textbf{A354}, 249 (2007).

\bibitem{helfer}Helfer, A. D., gr-qc/0008016. 

\bibitem{helfer3}Helfer, A. D., Int. J. Mod. Phys. \textbf{D13},
2299 (2004), gr-qc/0503052.

\bibitem{helfer2}Helfer, A. D., Rept.Prog.Phys. \textbf{66}, 943
(2003), gr-qc/0304042.

\bibitem{gibbons}Gibbons, G. W., in \emph{Proc. First Marcel Grossman
Meeting on General Relativity}, ed. R. Ruffini, 499 North-Holland
(1977).

\bibitem{jacobson}Jacobson, T., Phys. Rev. \textbf{D44}, 1731 (1991);
Phys. Rev. \textbf{D48}, 728 (1993); Prog The Phys Suppl \textbf{136},
1 (1999), hep-th/0001085.

\bibitem{jacobson2}Unruh, W. G., Phys. Rev., \textbf{D51}, 2827 (1995);
Brout, R., Massar, S., Parentani, R., and Spindel, Ph., Phys. Rev.
\textbf{D52}, 4559 (1995), hep-thy/9606121; Corely, S. and Jacobson,
T., Phys. Rev., \textbf{D53}, 6720, hep-th/9601973; Himemoto, Y. and
Tanaka, T., Phys. Rev. \textbf{D61}, 064004, gr-qc/9904076; Saida,
H. and Sakagami, M, Phys. Rev. \textbf{D61}, 084023, gr-qc/9905034.

\bibitem{messiah}Messiah, A.,\textit{ Quantum Mechanics}, North Holland,
(1961).

\bibitem{dewitt}Wipf, A., in \emph{Black Holes: Theory and Observation,
Proc. 179th W. E. Heraeus Seminar}, ed. F. W. Hehl, C. Kiefer, and
R. J. K. Metzler, 385, Springer (1998), hep-th/9801025; DeWitt, B.,
\textit{The Global Approach to Quantum Field Theory}, Oxford (2003).

\bibitem{fulling}Fulling, S. A., J. Phys. \textbf{A10}, 917 (1977);
Wald, R. M.,\emph{ Quantum Field in Curved Spacetime and Black Hole
Thermodynamics,} U. Chicago Press (1994).

\bibitem{hartle}Hartle, J. B., and Hawking, S. W., Phys. Rev. \textbf{D13},
2188 (1976).

\bibitem{birrel-fulling}Birrell, N. D., and Davies, P. C. W., \emph{Quantum
Fields in Curved Space}, Cambridge Univ. Press (1982); Fulling, S.
A., \emph{Aspects of Quantum Field Theory in Cureved Space-Time},
Cambridge Univ. Press (1989).
\end{thebibliography}
\end{document}